
\magnification=1200
\headline{\ifnum\pageno=1 \nopagenumbers
\else \hss\number \pageno \fi\hss}
\footline={\hfil}
\parindent=12pt
\baselineskip=20pt
\hsize=15 truecm
\vsize=23 truecm
\hoffset=0.7 truecm
\voffset=1 truecm
\overfullrule=0pt
\centerline{\bf HOW TO LIMIT RADIATIVE CORRECTIONS}\par
\centerline{\bf TO THE COSMOLOGICAL CONSTANT BY ${\bf M}^{\bf 4}_{\bf
Susy}$} \par
\vskip 5 truemm
\centerline{\bf Ulrich ELLWANGER} \par
\vskip 2 truemm
\centerline{Laboratoire de Physique Th\'eorique et Hautes
Energies\footnote{*}{Laboratoire
associ\'e au Centre National de la Recherche Scientifique - URA 63}}
\par
\centerline{Universit\'e de Paris XI, b\^atiment 211, 91405 Orsay Cedex,
France} \par

\vskip 3 truecm
\noindent \underbar{{\bf Abstract}} \par
Supergravity models are constructed in which the effective low energy
theory contains only ``super-soft''
explicit supersymmetry breaking: masses of the scalars and
pseudoscalars within a multiplet are split in opposite directions.
With this form of supersymmetry breaking the
radiative corrections of the matter sector to the vacuum energy are
bounded by ${\cal O} (M^4_{Susy})$ to all orders in
perturbation theory, and we require $Str \ M^2 = 0$ including the hidden
sector. The models are based on
K\"ahler potentials obtained in recent orbifold compactifications, and we
describe the construction of realistic theories.
 \par

\vskip 3 truecm
\noindent LPTHE Orsay 94-106 \par
\noindent December 1994
\vfill \supereject
The smallness of the cosmological constant belongs to the unsolved
problems in particle
physics. Within the standard model, we obtain a
vacuum energy and hence a
cosmological constant of ${\cal O} (M^4_{Weak} )$ [1]. \par

The situation becomes even worse, if we extend the validity of the
standard model (with the
help of fine tuning in order to keep $M_{Weak}$ stable) up to larger
scales as $M_{GUT}$. At
least quantum corrections then generate a vacuum energy of ${\cal
O}(M^4_{GUT})$. Supersymmetry, which helps to keep $M_{Weak}$ stable
under radiative
corrections without fine tuning, is also of some help concerning the
problem  of the
cosmological constant. Whereas unbroken global supersymmetry implies a
va\-ni\-shing vacuum
energy [2], soft explicit susy breaking (of ${\cal O} (M_{Susy}) \sim
{\cal O}(M_{Weak})$) leads to
non-vanishing radiative corrections to the vacuum energy. \par

The magnitude
of these contributions
depends on the type of the soft susy breaking terms. Using $M_{Planck}$
as an ultraviolet cutoff,
the typically assumed ``semi-soft'' gaugino masses, trilinear couplings
among the scalars and
identical masses for the scalar and pseudoscalar components of a chiral
multiplet generate a
vacuum energy of ${\cal O}(M^2_{Susy}\cdot M^2_{Planck})$. There exists
a further possibility to break
supersymmetry explicitely, but softly: if we denote the complex scalar
component
of a chiral multiplet
by $A$, a term of the form
$${m^2 \over 2} (A^2 + \overline{A}^2) \eqno(1)$$
\noindent gives opposite contributions to the masses squared of its real
and imaginary (or scalar and
pseudoscalar) components. After expressing the explicit susy breaking in
terms of a spurion
field in superspace [3] and using the corresponding power counting rules
it is
straightforward to show, that in this case the maximal radiative
contributions to the vacuum
energy, to all orders in perturbation theory, are bounded by ${\cal O}
(M^4_{Susy})$ (with $m^2
\sim M^2_{Susy})$ [4]. This is certainly a big step forward compared
with ${\cal
O}(M^2_{Susy} \cdot M^2_{Planck})$. The generally assumed ``semi-soft''
form of explicit susy
breaking thus does not exploit the potential power of supersymmetry to
bound the magnitude of
radiative corrections in the same way as a cutoff $\Lambda \sim
M_{Susy}$ would do.  \par

Subsequently we will consider supergravity theories which are motivated
by superstring models
(typically involving orbifold compactifications). Recently pro\-blems
related to the minimization
of the potential in the presence of contributions to the vacuum energy
of ${\cal O}(M_{Susy}^2 \cdot
M^2_{Planck} $) have been pointed out for this kind of theories [5].
\par

To one loop level, the absence of radiative contributions of ${\cal
O}(M^2_{Susy} \cdot
M^2_{Planck})$ to the Coleman-Weinberg effective potential [6] boils
down to the condition
of a vanishing $Str \ M^2$.
 General supergravity theories with $Str \ M^2=0$ have been discussed in
[7, 8].
In the presence of explicit supersymmetry breaking one needs, however, a
guiding principle like super-soft susy breaking; otherwise, higher loop
orders will inevitably
generate contributions of ${\cal O} (M^2_{Susy} \cdot M^2_{Planck} )$,
even if they happen to be
absent at the one loop level. \par

The questions arise, whether within string motivated theories such
models with only super-soft
susy breaking can be obtained, and whether these models can be
rea\-lis\-tic. Both questions will
be answered positively below. Actually, supergravity theories with
super-soft susy breaking
have already been developed and discussed in [4, 9]. These theories were
constructed
such that the minimization of the va\-cuum energy generates
automatically a hierarchy $M_{Susy}
\ll M_{Planck}$ (which, for self-consistency, requires the vacuum energy
to be of ${\cal
O}(M^4_{Susy}))$.
The resulting constraints on the parameters and particle spectrum are,
however, very tight
[10]. We will not discuss the generation of the hierarchy $M_{Susy}$ vs.
 $M_{Planck}$
here, but concentrate on general features of a vacuum energy of ${\cal
O}(M^4_{Susy})$. \par

The problem is complicated by the fact, that generally the light
particle content of these
theories is split into an ``observable'' matter sector and a ``hidden''
sector (typically the
graviton, dilaton and moduli superfields), whose interactions with the
matter sector are suppressed
by inverse powers of $M_{Planck}$. Only the matter sector makes up the
softly broken supersymmetric
theory discussed up to now, and higher loop corrections involving the
hidden sector are beyond the
scope of this paper. We will take the following attitude: due to the
weak interactions between
the matter and hidden sectors we require that each sector generates a
maximal contribution to
the vacuum energy of ${\cal O} (M^4_{Susy})$ separately. To one loop
level we are able to
control the hidden sector with the help of the vanishing of $Str \ M^2$;
within the matter sector with only super-soft supersymmetry breaking
$Str \ M^2$ vanishes automatically  [11]. To higher, and
actually arbitrary loop level we are able to control the matter sector
with the help of
the above-mentioned theorem on the vacuum energy in super-softly broken
susy. This way we
will obtain realistic theories, whose behaviour towards the vacuum
energy is as gentle as
possible within nowadays available technologies. \par

In order to construct the supergravity theory we have to specify the
K\"ahler potential $K$, the
superpotential $W$ and the gauge kinetic function $f$ [12]. The chiral
superfields within the
matter sector will be denoted by the letters $C_i$ and $A_i$; the total
number of $C_i$
\underbar{and} $A_i$ fields will be denoted by $N_c$. The hidden sector
contains a dilaton $S$
and $N_{M}$ moduli fields denoted by $T$, $M_i$ with $i = 1 \dots N_{M}
- 1$. The moduli
field $T$ is singled out to play the role of an overall ``breathing''
mode. The total number of
fields is thus given by $N = N_c + N_{M} + 1$. \par

The purpose of this paper is to propose a supergravity theory, which
leads to an effective low
energy theory with exlusively super-soft susy breaking. Its K\"ahler
potential is given by
$$K = - \ell n (S + \overline{S}) - 3 \ell n \left ( T + \overline{T} -
C_i \overline{C}_i - A_i \overline{A}_i \right ) $$
$$+ h (T, \overline{T}) \left ( A_i A_i + \overline{A}_i \overline{A}_i
\right ) + \tilde K(M_i , \overline{M}_i) \ \ \ . \eqno(2)$$
The first two terms are familiar from the construction of
four-dimensional superstring
theories [13], the second already from ``No-scale''  [14] or $SU(N, 1)$
[4, 9]
supergravity theories. The third term, involving the function $h(T,
\overline{T})$, has been
proposed as a solution of the so-called $\mu$-problem of the $MSSM$ in
[15].
Recently it has also been shown to appear in orbifold compactified
superstring theories
[16, 17]. In order to allow for the term of the form $A_iA_i + h.c.$,
the fields $A_i$ have to
transform as real representations under all internal symmetries. (Of
course, $A_iA_i + h.c.$ could be
replaced by $A_iB_i$ with $B_i$ transforming as the complex conjugate
representation of
$A_i$).
The part of the K\"ahler potential involving the moduli $M_i$,
$\tilde{K}(M_i ,
\overline{M}_i)$, is just required to lead to a positive definite metric
$\tilde{K},^{M_i}_{\phantom{M_i}\overline{M}_i}$. \par

For the
superpotential $W$ we make the ansatz
$$
W = \mu (S, M_i) + \tilde{W}(C_i, A_i) \eqno(3)
$$
\noindent where every term in $\tilde{W}$ is cubic in the fields $C_i$,
$A_i$. The first term $\mu$ is a
familiar result of gaugino condensation in the matter sector [18]. For
the gauge kinetic
function $f$ we also assume a field dependence of the form
$$
f(S, M_i) \ \ \ . \eqno(4)
$$
\noindent Below we will assume a form of the function $h(T,
\overline{T})$ such that all components of
the fields $A_i$ have positive masses squared and hence vanishing vevs.
With $<A_i> = 0$ one
finds the identity $({\cal G} = K + \ell n |W|^2)$
$$
{\cal G},_{T} ({\cal G},^T _{\phantom{T} \overline{T}})^{-1} {\cal
G},^{\overline{T}} = 3 \ \ \ . \eqno(5)
$$
\noindent The tree level scalar potential [12]
$$
V_{Tree} = e^{\cal G} \left ( {\cal G},_{I} ({\cal G},^I_{\phantom{I}
\overline{J}})^{-1} {\cal
G},^{\overline{J}} - 3 \right )  \eqno(6)$$
is then easily seen to be positive semi-definite, and minimized for
$$
{\cal G},_{S} = {\cal G},_{M_i} = {\cal G},_{C_i} = {\cal G},_{A_i} = 0
\ \ \ . \eqno(7)
$$
\noindent Thus one obtains a ``Goldstino angle'' [19] $\theta = 0$. The
vev of the field $T$ is
undetermined at this level, but in any case the tree level vacuum energy
vanishes exactly.
Supersymmetry is spontaneously broken; the corresponding gravitino mass
is given by (in the units
$M_{Planck}/8 \pi = 1$)
$$
m_{3/2} = e^{\cal G} = {\mu \ e^{\tilde{K}} \over (S + \overline{S})(T +
\overline{T})^3} \ \ \
. \eqno(8)$$

In order to derive the effective low energy theory for the matter sector
it is of considerable
help to note that among all possible ${\cal G},_{I}$ only ${\cal
G},_{T}$ is nonzero. From ${\cal
G},_{S} = {\cal G},_{ M_i} = 0$ it follows, e.g., that no susy breaking
gaugino masses are
present at tree level. From an investigation of the scalar and fermionic
interactions, and an
appropriate rescaling of the fields in order to achieve canonical
kinetic energies, one finds
that the effective low energy theory is described by a superpotential
$W_{eff}$ of the form
$$
W_{eff} = {1 \over (S + \overline{S})^{1/2}} \tilde{W}(C_i, A_i) + {M_A
\over 2} A_i A_i \eqno(9)
$$
\noindent with
$$
M_A = 2 m_{3/2} (T + \overline{T}) \left ( h + (T + \overline{T})
h,_{\overline{T}} \right ) \ \
\ . \eqno(10)$$
\noindent The only susy breaking interactions are indeed of the form of
eq. (1), with
$$
m^2 = - 2 m_{3/2}^2 (T + \overline{T})^2 \left (  h,_{T} + h,_{
\overline{T}} + (T +
\overline{T}) h,_{T\overline{T}} \right  ) \ \ \ .  \eqno(11)$$
\noindent Thus we have accomplished the first part of our task. Next we
turn to the condition of the
vanishing supertrace $Str \ M^2$, in order to tame one loop
contributions to the vacuum energy
of ${\cal O}(M_{Susy}^2 \cdot M_{Planck}^2)$ including the hidden
sector. In the present case of
vanishing gaugino masses, the formula of [20] (see also [8]) for $Str \
M^2$ for general
supergravity theories boils down to
$$
Str \ M^2 = 2m_{3/2}^2 \left [ N - 1 - {\cal G},^I \ \partial_I \
\partial_{\overline{J}} \ \ell n
\det ({\cal G},_{M \overline{N}}) {\cal G},^{\overline{J}} \right ] \ \
\ . \eqno(12)$$
\noindent In our case, with ${\cal G},^I = 0$ except for $I = T$, we
only need
$$
{\cal G},^T \ \partial_T \ \partial_{\overline{T}} \ \ell n \det ({\cal
G},_{ M \overline{N}}) {\cal
G},^{\overline{T}} = 2 + N_c + {\cal O}(A, \overline{A}) \ \ \ .
\eqno(13)
$$
\noindent With $<A> = 0$ and $N = N_c + N_{M} + 1$ we thus obtain
$$
Str \ M^2 = 2m_{3/2}^2 \left [ N_{M} - 2 \right ] \ \ \ , \eqno(14)
$$
\noindent and the condition of a vanishing supertrace just becomes $N_M
= 2$. This way we have
constructed a model of the class discussed in [8]; we have obtained,
however, a model with a
special property: since, within the observable sector, susy is only
broken by mass terms of the
form of eq. (1), the part of the supertrace to which the observable
sector contributes vanishes by
itself [11]. Hence, with $N_M = 2$, the contribution of the hidden
sector to the supertrace vanishes
by itself as well. Thus we have satisfied all necessary conditions,
which are required  in order that the contributions
to the vacuum energy are limited by ${\cal O}(M_{Susy}^4$) from the
observable sector to all orders in
perturbation theory, and those from the hidden sector to one loop order.
\par

Let us now discuss the possibility of constructing realistic models
within this class of
theories. Apart from a reasonable particle content we require
sufficiently large masses for the
gauginos, squarks and sleptons, and negative masses squared in the Higgs
sector in order to
trigger $SU(2) \times U(1)$ symmetry breaking. Generally this is
achieved, if not already at
tree level, with the help of radiative corrections. In the presence of
semi-hard soft terms, the
most important radiative corrections are logarithmically divergent and
most conveniently summed
up by integrating the renormalization group equations from the cutoff
scale $M_{Planck}$ down to
the weak scale or $M_{Susy}$. \par

In a model with exclusive super-soft susy breaking, however, the only
logarithmically divergent
contributions to the effective action (apart from wave function
normalizations) are proportional
to $F$-components of singlet superfields [3]. Consequently, the right
hand sides of the RG
equations for gaugino, squark and slepton masses as well as trilinear
scalar couplings vanish.
Nevertheless such masses are radiatively generated; now, however, the
corresponding Feynman
diagrams are ultraviolet finite. \par

Let us first have a look at the radiative corrections involving gauge
interactions. We assume that
the fields $A_i$, with masses $M_A$ and mass splittings as in eq. (1),
transform as representations
$r$ under gauge groups $a$. For $m^2 \ll M_A^2$ the corresponding
gauginos receive one loop masses
given by
$$m_a = \sum_r {\alpha_a m^2 \over 4 \pi M_A} T_r^a \ \ \ . \eqno(15)$$
\noindent The Casimir eigenvalue $T_r^a$ has to be replaced by the
charge squared in the case of
a $U(1)$ gauge group, and is given by 1/2 resp. $N$ for $r$ denoting a
fundamental resp. adjoint
representation of $SU(N)$. At two loop order, all scalars of the theory
which transform as
representations $q$ under the gauge groups $a$ obtain positive definite
masses squared given by
[21]
$$m_q^2 = \sum_{a,r} {\alpha_a^2 m^4 \over 8 \pi^2 M_A^2} T_r^a C_q^a \
\ \ . \eqno(16)$$
\noindent In the case of $U(1)$ $C_q^a$ is again given by the square of
the charge of $q$, whereas
$C_q^a = {N^2 - 1 \over 2N}$ resp. $N$ for fundamental resp. adjoint
representations of $SU(N)$.
\par

The most important Yukawa mediated radiative corrections are the
above-mentioned logarithmically
divergent ones, if a singlet field couples to the massive fields $A$.
Let us assume a corresponding
term in the superpotential $W_{eff}$,
$$W_{eff} = \beta S AA + W' \ \ \ . \eqno(17)$$
\noindent Then, to one loop order, the following contribution to the
effective potential is
obtained:
$$- {\beta \over 16 \pi^2} m^2 \left ( F_S + \overline{F}_S \right )
\ell n {\Lambda^2 \over M_A^2}
\eqno(18)$$
\noindent where $F_S$ is an expression quadratic in the scalar fields
given by $F_S = {\partial
W_{eff} \over \partial S}$. The term (18) is actually the presently
harmless result of the singlet
tadpole contribution, which can mess up hierarchies in the presence of
semi-soft susy breakings
[22]. Further Yukawa induced radiative corrections are generally
negligible for $m^2 \ll M_A^2$~;
the interaction in eq. (17), e.g., gives additionally rise to a positive
mass $m_S^2$ for $S$ of
$$m_S^2 = {\beta^2 \over 8 \pi^2} \ {m^6 \over M_A^4} \ \ \ .
\eqno(19)$$
\noindent and a term linear in $S$
$${\beta \over 16 \pi^2} \ {m^4 \over M_A} (S + \overline{S}) \ \ \ .
\eqno(20)$$
\par

These results can already be used to discuss the construction of
realistic models. First,
in order to generate realistic gaugino masses for all gauginos of the
standard model gauge groups
via (15), the fields $A_i$ should carry quantum numbers under all these
gauge groups. Second, in
order for gaugino, squark and slepton masses via (15) and (16) to be
sufficiently large, the
ratio ${m^2 \over M_A}$ should be at least of ${\cal O}(TeV)$. Both
arguments rule out the
attractive possibility of identifying $AA$ with the $MSSM$ Higgs fields
$H_1H_2$ as in [23] (and
choosing $m^2 > M_A^2$ in order to generate Higgs VEVs already at tree
level). \par

Instead, the fields $A$ have to be identified with new fields beyond the
MSSM, with $M_A >
{\cal O}(TeV)$. The successful unification of the running gauge
couplings at $M_{GUT}$ within
the $MSSM$ is not spoiled, if the fields $A$ are chosen to fill up
complete representations of
$SU(5)$ as, e.g., 5 + $\overline{5}$. \par

A realistic class of models can thus be constructed as follows: we
identify the fields $C_i$
of eqs. (2) and (3) with the quark, lepton and Higgs superfields of the
$MSSM$ as well as a gauge
singlet superfield $S$. In addition the matter sector contains the
fields $A_i$. For the cubic part $\tilde{W}$ of the
superpotential in eq. (3) we choose
$$\tilde{W} = \tilde{\lambda} S H_1 H_2 + {\tilde{k} \over 3} S^3 +
\tilde{\beta} S A_i A_i + \dots
\eqno(21)$$
\noindent where the dots denote the $MSSM$ Yukawa couplings. According
to eq. (9) the effective
superpotential $W_{eff}$ then reads
$$W_{eff} = \lambda S H_1 H_2 + {k \over 3} S^3 + \beta S A_i A_i + {M_A
\over 2} A_i A_i + \dots
\eqno(22)$$
\noindent with rescaled Yukawa couplings and $M_A$ given by eq. (10).
Note that for generic
functions $h(T, \overline{T})$ and vevs of $T$ of ${\cal O}(M_{Planck})$
we have from eq. (11)
$$M_A \sim m \sim M_{Susy} \ \ \ ; \eqno(23)$$
\noindent the condition $M_A > m$ turns into a condition on the form of
$h(\overline{T}, T)$ and
the vev of $T$. ($M_A \gg m$ just simplifies the computation of the
radiative corrections.) \par

Now, at one resp. two loop order, gaugino as well as positive squark,
slepton and Higgs masses are
generated according to eqs. (15) and (16). The logarithmically divergent
contribution (18), with
$F_S = \lambda H_1H_2 + kS^2 + \beta A_iA_i$, generates Higgs masses of
the form
$$m_3^2 H_1H_2 + h.c. \eqno(24a)$$
\noindent with
$$m_3^2 = - {\lambda \beta d_r \over 16 \pi^2} m^2 \ell n {\Lambda^2
\over M_A^2} \ \ \ ,
\eqno(24b)$$

\noindent which can destabilize the Higgs potential as desired. ($d_r$
denotes the dimension of the
representation $r$ of the fields $A_i$). Actually, the positive two loop
stop masses $m_{st}^2$ from
eq. (16) induce, at three loop order, negative masses squared for the
Higgs fields $H_2$, which have
Yukawa couplings $h_t$ to the top quarks [24]. For $m_{st}^2 \ll m^2$
this term can be estimated to
be
$$- {3 \over 4 \pi^2} h_t^2 m_{st}^2 |H_2|^2 \ell n {M_A \over m_{st}} \
\ \ . \eqno(25)$$
\noindent With $m_{st}^2$ given by eq. (16), where $\alpha_{QCD}$ gives
the leading
contribution, this negative mass for $H_2$ can be numerically larger
than the positive mass
for $H_2$ obtained via eq. (16), where only the electroweak gauge
couplings appear. Thus
there exist even two possible mechanisms to trigger the desired $SU(2)
\times U(1)$ symmetry
breaking.  \par

We have checked that, after minimization of the complete scalar
potential in\-clu\-ding the $S$
dependent terms (19) and (20), a wide range of parameters $\lambda$,
$k$, $\beta$, $M_A$ and
$m$ exists, which leads to realistic particle masses. The details of the
spectrum depend, in
addition, on the representation $r$ of the fields $A_i$. A complete
analysis of the full
range of parameters is beyond the scope of the present paper. Some
limiting situations
are, however, worth mentioning: one may choose the Yukawa coupling $k$
vanishingly small
or even equal to zero. Then it is only the mass $m_S^2$ of eq. (19),
which stabilizes the
potential for $S$, and $S$ assumes a large vev of ${\cal O}(M_A^3/\beta
m^2)$. Accordingly
the ratio $M_A/m$ has not to be chosen too large, and $\lambda$ has to
be small in order
to allow for vevs of $H_1$ and $H_2$. The spectrum will contain a light
pseudoscalar
(dominantly gauge singlet), since the Peccei-Quinn $U(1)$ symmetry in
the $H_1$, $H_2$, $S$ - sector
is only weakly broken by the terms (18), (20). \par

Independently thereof, if the $F$ term (24) plays the dominant role in
triggering
nonvanishing vevs of $H_1$ and $H_2$, $\tan \beta = {<H_2> \over <H_1>}$
will be close to
1, whereas $\tan \beta$ will be large in the case where the
$h_t$-induced contribution
(25) is most important. \par

Let us conclude with some comments on the prospects of realizing the
ansatz for $K$
(eq. (2)), $W$ (eq. (3)) and $f$ (eq. (4)) within string theory. The
first two terms in
eq. (2) and the form of $f$ in eq. (4) are well known from string theory
at tree level.
Also a function $h(T, \overline{T})$ has been found to arise at tree
level [16, 17]; the
corresponding expression is of the form
$$h(T, \overline{T}) \sim {1 \over T + \overline{T}} \ \ \ . \eqno(26)$$
\noindent With such a function $h(T, \overline{T})$ one finds, from eqs.
(10) and (11),
$M_A = m = 0$. However, string loop corrections generate functions $h(T,
\overline{T})$
different from (26) [17]. Generally they also modify the first two terms
of eq. (2)
[25], and the ansatz for $f$ eq. (4) [26]. The details depend, of
course, on the string
model under consideration (as, e.g., on the Green Schwarz angles
$\delta_{GS}^i$). For the present
class of models to emerge it would thus be desirable that string loop
corrections leave the
K\"ahler potential $K$ and the gauge kinetic function $f$ untouched, but
modify the form
of the function $h(T, \overline{T})$. It remains to be seen, however,
whether such
string models can be constructed. \par

The purpose of this paper is to point out the existence of supergravity
theories with exclusive
supersoft supersymmetry breaking, in which even radiative corrections
generate a vacuum energy of not
more than ${\cal O}(M_{Susy}^4)$. We emphasize the possibility of
constructing realistic models
within these theories along the lines discussed above. One could imagine
that string
theory would like to make use of this attractive possibility to control
the quantum
corrections to the cosmological constant. This would lead to important
restrictions
on the string models resp. string vacua.

 \vfill \supereject
\centerline{\bf \underbar{References}} \par \bigskip
\item{[1]} A. Linde, JETP Letters \underbar{19} (1974) 183.
\item{[2]} B. Zumino, Nucl. Phys. \underbar{B89} (1975) 535.
\item{[3]} L. Girardello, M. Grisaru, Nucl. Phys. \underbar{B194} (1982)
65.
\item{[4]} U. Ellwanger, N. Dragon, M. Schmidt, Nucl. Phys.
\underbar{B255} (1985) 549.
\item{[5]} K. Choi, J. E. Kim, H. P. Nilles, Phys. Rev. Lett.
\underbar{73} (1994) 1758.
\item{[6]} S. Coleman, E. Weinberg, Phys. Rev. \underbar{D7} (1973)
1888.
\item{[7]} S. Ferrara, C. Kounnas, M. Porrati, F. Zwirner, Phys. Lett.
\underbar{194B} (1987) 366,
\item{} C. Kounnas, M. Quiros, F. Zwirner, Nucl. Phys. \underbar{B302}
(1988) 403.
\item{[8]} S. Ferrara, C. Kounnas, F. Zwirner, CERN preprint TH-7192/94.
\item{[9]} N. Dragon, U. Ellwanger, M. Schmidt, Phys. Lett.
\underbar{145B} (1984) 192, Phys. Lett.
\underbar{154B} (1985) 373, Z. Phys. \underbar{C29} (1985) 219, Progr.
in Part. and Nucl. Phys.
\underbar{18} (1987) 1.
\item{[10]} U. Ellwanger, Z. Phys. \underbar{C47} (1990) 281.
\item{[11]} S. Ferrara, L. Girardello, F. Palumbo, Phys. Rev.
\underbar{D20} (1979) 403.
\item{[12]} E. Cremmer, S. Ferrara, L. Girardello, A. Van Proyen, Nucl.
Phys. \underbar{B212} (1983) 413.
\item{[13]} E. Witten, Phys. Lett. \underbar{B155} (1985) 151.
\item{[14]} E. Cremmer, S. Ferrara, C. Kounnas, D. Nanopoulos, Phys.
Lett. \underbar{B133} (1983)
61; A. Lahanas, D. Nanopoulos, Phys. Rep. \underbar{145} (1987) 1.
\item{[15]} G. Giudice and A. Masiero, Phys. Lett. \underbar{B206}
(1988) 480.
\item{[16]} G. Lopes Cardoso, D. L\"ust, T. Mohaupt, Nucl. Phys.
\underbar {B432}  (1994) 68.
\item{[17]} I. Antoniadis, E. Gava, K. Narain, T. Taylor, Nucl. Phys.
\underbar{B432} (1994) 187.
\item{[18]} J.-P. Derendinger, L. Ib\'a$\tilde{\rm n}$ez, H.P. Nilles,
Phys. Lett. \underbar{B155} (1985) 65;
\item{} M. Dine, R. Rohm, N. Seiberg, E. Witten, Phys. Lett.
\underbar{B156} (1985) 55.
\item{[19]} A. Brignole, L. Ib\'a$\tilde{\rm n}$ez, C. Mu$\tilde{\rm
n}$oz, Nucl. Phys. \underbar{B422} (1994) 125.
\item{[20]} M. Grisaru, M. Rocek, A. Karlhede, Phys. Lett.
\underbar{B120} (1983) 110.
\vfill \supereject
\item{[21]} M. Dine, W. Fischler, Nucl. Phys. \underbar{B204} (1982)
346;
\item{} L. Alvarez-Gaum\'e, M. Claudson, M. Wise, Nucl. Phys.
\underbar{B207} (1982) 96.
\item{[22]} J. Polchinski, L. Susskind, Phys. Rev. \underbar{D26} (1982)
3661;
\item{} H.-P. Nilles, M. Srednicki, D. Wyler, Phys. Lett.
\underbar{B124} (1983) 337;
\item{} A. Lahanas, Phys. Lett. \underbar{B124} (1983) 341;
\item{} U. Ellwanger, Phys. Lett. \underbar{B133} (1983) 187;
\item{} J. Bagger and E. Poppitz, Phys. Rev. Lett. \underbar{71} (1993)
2380.
\item{[23]} A. Brignole, F. Zwirner, preprint CERN-TH.7439/94, hep-th
9409099.
\item{[24]} L. Ib\'a$\tilde{\rm n}$ez, G. G. Ross, Phys. Lett.
\underbar{110B} (1982) 215.
\item{[25]} J.-P. Derendinger, S. Ferrara, C. Kounnas, F. Zwirner, Phys.
Lett. \underbar{B271}
(1991) 307 and Nucl. Phys. \underbar{B372} (1992) 145;
\item{} I. Antoniadis, E. Gava, K. Narain, T. Taylor, Nucl. Phys.
\underbar{B407} (1993) 706.
\item{[26]} L. Ib\'a$\tilde{\rm n}$ez, H.-P. Nilles, Phys. Lett.
\underbar{B169} (1986) 354;
\item{} H. Itoyama, J. Leon, Phys. Rev. Lett. \underbar{56} (1986) 2352;
\item{} L. Dixon, V. Kaplunovsky, J. Louis, Nucl. Phys. \underbar{B255}
(1991) 649.

 \bye